\documentstyle[12pt]{article} 
\font\tenrm=cmr10 

\newcommand{\bref}[1]{(\ref{#1})} 
\newcommand{\ct}[1]{\cite{#1}}

\newcommand{\be}{\begin{equation}} 
\newcommand{\ee}{\end{equation}} 

\def\theequation{\thesection.\arabic{equation}}
\def\@eqnnum{{\rm (\theequation)}}

\def\lsim{\mathrel{\rlap{\lower4pt\hbox{\hskip1pt$\sim$}}
    \raise1pt\hbox{$<$}}}
\def\gsim{\mathrel{\rlap{\lower4pt\hbox{\hskip1pt$\sim$}}
    \raise1pt\hbox{$>$}}}
\def\frac#1#2{{{#1} \over{#2}}} 

\begin{document}  
\begin{titlepage} 
  
\begin{flushright}  
{CCNY-HEP-96/7 \\ }   
{hep-lat/9612004 \\ }  
{\hfill May 1996 \\ }  
\end{flushright}  
\vglue 1cm  
	   
\begin{center}   
{ 
{\Large \bf A Strong-Coupling Analysis \\ 
of the Lattice $CP^{N-1}$ Models \\  
in the Presence of a $\theta$ Term \\ }  
\vglue 1.0cm  
{Jan C.\ Plef\/ka $^{1}$ \\ }   
\vglue 0.25cm  
{and} 
\vglue 0.25cm  
{Stuart Samuel $^{2}$ \\ }   
\vglue 1.0cm  

{\it Department of Physics\\}
{\it City College of New York\\}
{\it New York, NY 10031, USA\\} 
\vglue 0.8cm  
 
\vglue 0.3cm  
  
{\bf Abstract} 
} 
\end{center}  
{\rightskip=3pc\leftskip=3pc 
\quad A $\theta$ term, 
which couples to 
topological charge, 
is added to the lattice 
$CP^{N-1}$ model.  
The strong-coupling character expansion 
is developed. 
The series for the free energy and mass gap are respectively 
computed to tenth order and fourth order.  
Several features of the strong-coupling analysis emerge.  
One is the loss 
of superconfinement.  
Another is that in the intermediate coupling constant region, 
there are indications of a transition 
to a deconfining phase when $\theta$ is sufficiently large. 
The transition is like the one which has been observed 
in Monte Carlo simulations 
of a similar lattice $CP^{N-1}$ action.  

}

\vfill

\textwidth 6.5truein
\hrule width 5.cm
\vskip 0.3truecm 
{\tenrm{
\noindent 
$^{1)}$ \hspace*{0.2cm}E-mail address: plefka@scisun.sci.ccny.cuny.edu \\ }
$^{2)}$\hspace*{0.35cm}E-mail address: samuel@scisun.sci.ccny.cuny.edu \\ }
 
\eject 
\end{titlepage} 

\newpage  

\baselineskip=20pt

{\bf\large\noindent I.\ Introduction}\vglue 0.2cm
\setcounter{section}{1}   
\setcounter{equation}{0}   

One of the unresolved puzzles of the standard model 
is the strong CP problem.  
The problem emerged after the discovery of instantons in 
four-dimensional Yang-Mills theories 
\ct{bpst75}.  
Instantons represent barrier penetration processes 
between different classical $n$-vacua.  
The ``true'' quantum vacua are believed to be linear 
combinations of $n$-vacua weighted by a phase 
$\exp \left( { i n \theta } \right)$ 
\ct{cdg76,jr76}.  
Alternatively, 
one can add $\theta \int d^{4} x \nu (x)$, 
where 
$ \nu (x) = g^2 F^a \tilde F^a (x) / (32 \pi^2) $, 
to the action lagrangian.  
Here, $\int d^{4} x \nu (x) = Q$ 
is the topological charge.  
It is one (minus one) for an instanton (anti-instanton).  
The action term 
$\theta \int d^{4} x  \nu (x)$ 
breaks parity, 
time-reversal invariance 
and CP symmetry.  
Although the $\theta$-term is a total derivative  
so that it does not contribute in perturbative Feynman diagrams, 
instantons and topological quantum fluctuations 
contribute to it. 
Such topological fluctuations are known 
to exist because they contribute significantly 
to the $\eta^\prime$ mass 
\ct{fgl73}--\ct{veneziano79}.  
Lattice simulations have computed 
these topological fluctuations 
and yield results 
of the right magnitude   
to explain $m_{\eta^\prime}$ 
\ct{woit83}--\ct{cdpv90}.  

When quarks are included, 
phases in the quark mass matrix ${\cal M}$ 
contribute to CP violation also.  
Using a $U(1)$ axial rotation, 
a common phase in ${\cal M}$ can be shifted 
to the $\theta$-term  
due to the axial anomaly 
\ct{adler69,bj69}.  
The physical effective theta parameter $\theta_{eff}$ is 
$\theta_{eff} = \theta + Arg Det {\cal M}$.  
A non-zero value of $\theta_{eff}$ 
contributes, among things, 
to the electric dipole moment of the neutron.  
Requiring compatibility with experiment leads to 
$\theta_{eff} \lsim 10^{-9}$ 
\ct{baluni79,cdvw79}.  
The difficulty in understanding how $\theta_{eff}$ 
can naturally be so small 
constitutes the strong CP problem.  
  
Various solutions to the strong CP problem 
have been proposed 
\ct{kim87,cheng88,mohapatra92}.  
Perhaps the most theoretically attractive suggestion 
is the Peccei-Quinn mechanism 
\ct{pq77}.  
It predicts the existence of an axion 
\ct{weinberg78,wilczek78}. 
Such an axion has not been seen in accelerator experiments.  
A modification of the mechanism leads to an invisible 
axion 
\ct{kim79}--\ct{dfs81}.  
Astrophysical considerations and accelerator experiments 
have left open a somewhat narrow mass window for the invisible axion 
\ct{turner90}. 

Another possible solution is 
that the current-algebra mass of the up quark is zero 
\ct{gm81,km86,cks88}.  
Various analyses suggest that the up-quark mass is non-zero 
and equal to about $5$ MeV 
\ct{weinberg77,gl82}.  
However, 
the up-quark mass has been the subject of much debate 
and some theorists feel that a zero value is not ruled out   
\ct{bsn94}.  

The Peccei-Quinn mechanism involves a relaxation field  
which allows $\theta_{eff}$ to be rotated away.  
This field is the axion.  
In ref.\ \ct{samuel92}, 
a $2+1$ dimensional model was considered 
in which the Yang-Mills sector generated 
its own relaxation field.  
This led to a natural solution to the strong CP problem 
within the pure Yang-Mills sector.  
Criteria were established 
to determine when such a natural relaxation mechanism arises 
\ct{samuel92}. 
Unfortunately, 
it has not been possible to determine whether or not 
these criteria are satisfied 
for four-dimensional Yang-Mills theories.  

The strong CP problem in a pure Yang-Mills theory 
involves vacuum physics.  
Vacuum physics is governed by the long-distance behavior 
of a theory.  
The long-distance behavior of non-abelian gauge theories 
involves non-perturbative and strong-coupling effects.  
Since this regime 
is not so theoretically well understood, 
there is the possibility that some subtle mechanism 
exists which resolves the strong CP problem naturally.   

Given the complexity of the strong CP problem, 
it is useful to consider simpler systems.  
One class of such systems is 
the two-dimensional $CP^{N-1}$ models.   
These models are asymptotically free 
and have instantons for all $N \ge 2$.  
In addition, a $\theta$-term can be added 
to the action.  
The analysis in the large $N$ limit 
leads to considerable insight into their physics 
\ct{dlv78,witten79b}.  
The $\theta$ dependence is expected to be a $1/N$ effect -- 
this contrasts the exponentially suppressed behavior $\exp (-cN)$ 
expected from an instanton analysis 
\ct{witten79b}. 
In the leading order $1/N$ result, 
the $CP^{N-1}$ models consist of $N$ free particles 
and $N$ free antiparticles,   
and there is no $\theta$ dependence. 
When the first $1/N$ correction is taken into account, 
quantum fluctuations produce a dynamically generated $U(1)$ 
gauge field which, in turn, 
generates a constant force between 
particles and antiparticles, 
thereby leading to confinement.  
The free energy ${\cal F}$ per unit volume 
behaves as $ c \theta^2 / N$, where $c$ is a constant.  
It is expected that the confining force becomes stronger 
and stronger as higher-order $1/N$ corrections are included 
since the $CP^{N-1}$ models exhibit superconfinement 
\ct{samuel83}. 
In a theory with superconfinement, 
the confining force is so strong 
that an antiparticle must be positioned on a particle. 
Not only is there confinement at asymptotic distances 
but there is local confinement as well -- 
particles and antiparticles in a bound state cannot be 
separated at all.  
The dramatic changes in behavior 
in going from leading order to next-to-leading order 
are indicative of the singular nature of the $1/N$ expansion 
in the $CP^{N-1}$ models.  
The singular nature is also reflected 
in the fact that strong coupling and $1/N$ expansions 
do not commute 
\ct{samuel83}.  

G. Schierholz and co-workers 
have performed Monte Carlo simulations 
on the two-dimensional $CP^{3}$ model 
with a $\theta$ term 
\ct{schierholz94}. 
A priori surprising results have emerged.  
For fixed inverse coupling $\beta$, 
a phase transition occurs as a function of $\theta$.  
Letting $\theta_c$ denote the critical value, 
the free energy per unit volume is well represented by 
\be
  {\cal F} =
    \left\{ \begin{array}{ll} 
    a\left( \beta  \right) \theta^2 
       \quad &\theta \le \theta_c \\ 
    c\left( \beta  \right) \quad \, \quad &\theta \ge \theta_c   
            \end{array} 
     \right. 
\quad . 
\label{eq1} 
\ee
In other words, 
for $\theta \ge \theta_c$, 
the free energy has no dependence on $\theta$.  
In two dimensions, 
this implies that 
the string tension vanishes for $\theta \ge \theta_c$ 
so that confinement is lost.  
Furthermore, 
$\theta_c$ appears to go to zero as 
the coupling $g$ goes to zero 
(or $\beta$ goes to infinity). 
The behavior of ${\cal F}$ for $\theta \le \theta_c$ 
is compatible with the next-to-leading order $1/N$ result.  
This is not the case for $\theta > \theta_c$.  
Since additional $1/N$ corrections have not been computed, 
it is unclear whether there is a discrepancy.  

The two-dimensional Schwinger model 
has a cusp in the free energy at $\theta = \pi$, 
signaling the spontaneous breaking 
of CP invariance 
\ct{coleman76}.  
A similar phenomenon is expected in the $CP^{N-1}$ models.  
In fact, 
at infinite coupling, $\beta =0$, 
the lattice $CP^{N-1}$ models do undergo spontaneous CP breaking 
at $\theta = \pi$ 
\ct{seiberg84}.  
Hence, a phase transition with $\theta_c = \pi$ 
is not unexpected.  
The surprising feature of the $CP^{3}$ model 
is the dependence of $\theta_c$ on $\beta$ 
\ct{schierholz94}.  
It suggests that to obtain a confining theory 
from the continuum limit of a lattice $CP^{N-1}$ model, 
one must take $\theta$ to zero.  
Thus it appears that continuum confining $CP^{N-1}$ models 
must have $\theta = 0$.  
If the analog statement were true 
for a four-dimensional Yang-Mills theory, 
then the strong CP problem would be solved.  

An analysis 
\ct{schierholz95} 
of the free energy 
of the four-dimensional $SU(2)$ Yang-Mills theory  
indicates that the free energy behaves as in equation 
\bref{eq1}.  
In contrast to two-dimensional systems, however, 
a constant ${\cal F}$ behavior for $\theta \ge \theta_c$ 
does not necessarily indicate a loss of confinement.  
Nonetheless, if $\theta$ must be selected 
to be less than $\theta_c$ 
for certain as-of-yet-unknown physical reasons
and $\theta_c$ goes to zero as $\beta \to \infty$, 
then the strong CP problem would be solved.  
Up to now, 
simulations of the four-dimensional $SU(2)$ gauge theory 
have been done at only one value of $\beta$, 
so that it is not known whether 
$\theta_c$ goes to zero as $\beta \to \infty$. 

Since the strong CP problem is related to long-distance physics 
and since long distances are equivalent to strong couplings 
in an asymptotically free theory, 
it is of interest to perform lattice strong-coupling expansions.  
As far as we know, 
only one strong-coupling expansion 
has been performed in the presence of a $\theta$ term.  
The zeroth order and order $\beta^2$ terms 
for the free energy and mass gap 
were obtained in 
ref.\ \ct{seiberg84}.  
Strong-coupling series  
are best organized 
in character expansions.  
Character expansions were introduced 
in ref.\ \ct{migdal76} 
for lattice gauge theories.  
They were obtained for matrix-model spin systems 
and for the large $N$ limit   
in \ct{gs82}.  
Finally, 
the adaptation of such methods to vector-like spin systems, 
including the $CP^{N-1}$ models,  
was achieved 
in \ct{stone79,rs81}.  
As far as we know, 
the techniques for performing 
character expansions 
in the presence of a $\theta$ term 
have not yet been developed.  
One of the purposes of the current work 
is to fill this need.  
Another purpose is to gain insight 
from such analytic strong-coupling computations 
into the physics associated with $\theta$. 
Strong-coupling analysis 
is often one of the best ways to gain an understanding  
into the non-perturbative long-distance 
behavior of asymptotically free theories.  
It was by these means, for example, 
that K.\ Wilson demonstrated 
that non-abelian gauge theories have the potential 
to confine quarks 
in four dimensions 
\ct{wilson74}.

\bigskip 
{\bf\large \noindent II.\ Methodology}  
\setcounter{section}{2}   
\setcounter{equation}{0}   

The action for the continuum $CP^{N-1}$ model is 
\be
    S = \beta N \int d^d x  
 \left( {  \partial_\mu z_i^\ast \partial^\mu z^i + 
      \left( {  z_i^\ast \partial_\mu z^i } \right)
      \left( {  z_j^\ast \partial^\mu z^j } \right) 
  } \right) 
\quad , 
\label{actiona}
\ee 
where $z(x)$ is an $N$-component complex field satisfying 
$z_i^{\ast} z^i (x) = 1$.  
Eq.\ \bref{actiona} possesses the local symmetry 
$z(x) \to \exp ( i \alpha (x) ) z(x)$.  
Hence, the local gauge-invariant degrees of freedom 
at a point $x$ 
live on the manifold $CP^{N-1}$.  
Using an auxiliary field $A_\mu (x)$, 
the action in Eq.\ \bref{actiona} can be written as 
\be 
  S = \beta N \int d^d x  
    \left( { \partial_\mu - i A_\mu  } \right) z_i^\ast 
  \ \left( { \partial^\mu + i A^\mu  } \right) z^i  
\label{actionb}
\quad .  
\ee 
Quantum fluctuations turn $A_\mu (x)$ into a propagating field.  
It becomes the $U(1)$ field mentioned in the Introduction 
and confines the charged $+1$ particles $z_i^\ast$ 
to the charged $-1$ particles $z_i$.  

There are always many equally satisfactory ways to latticize 
a continuum action.  
For the $CP^{N-1}$ models, 
two lattice actions naturally arise --  
they correspond to the naive discretization 
of Eqs.\ \bref{actiona} and \bref{actionb}.  
A lattice version of \bref{actiona} is 
\be 
 S = \beta N \sum_{x,\Delta} 
  z_{x+\Delta}^\ast \cdot z_{x} \  z_{x}^\ast \cdot z_{x+\Delta} 
\quad , 
\label{lactiona}
\ee
while a lattice version of \bref{actionb} is 
\be 
  S = \beta N \sum_{x,\Delta} 
 \left( { 
    z_{x}^\ast \cdot z_{x+\Delta} 
    \exp \left( { iA_{x,x+\Delta} } \right) + 
    z_{x} \cdot z_{x+\Delta}^\ast 
    \exp \left( { -iA_{x,x+\Delta} } \right) 
 } \right) 
\quad .
\label{lactionb} 
\ee
In Eqs.\ \bref{lactiona} and \bref{lactionb}, 
the sum over $\Delta$ 
involves the $d$ positively-directed nearest neighbors to $x$,  
so that $\Delta$ takes on the values 
$e_1$, $e_2$,$\dots$, $e_d$, 
where $e_i$ is a unit vector in the $i$th direction.   
In what follows, 
we restrict to the case of two dimensions so that $d=2$. 

For the action 
in \bref{lactiona}, 
a $\theta$ term is defined as follows 
\ct{bl81}.  
Associate a $U(1)$ field 
$U \left({  x, x+\Delta } \right)$ with 
each link via 
$
 U \left({  x, x+\Delta } \right) = 
  z_{x}^\ast (x) \cdot z_{x+\Delta} / 
  \vert { z_{x}^\ast (x) \cdot z_{x+\Delta} } \vert
$.  
Let $U_p$ be 
the product of the $U \left({  x, x+\Delta } \right)$ 
around a plaquette.  
Define the local topological density $\nu_p $ via  
$\nu_p \equiv log \left( { U_p } \right) / (2\pi)$.  
The branch of the logarithm is taken 
to be between $-\pi$ and $\pi$. 
The total topological charge $Q$ is 
$Q = \sum_p \nu_p$.  
The $\theta$ term to be added to the action is $i \theta Q$,  
that is,  
\be 
  S_{\theta \ {\rm term}} = 
  {{ i\theta } \over {2\pi} } \sum_p 
   log \left( { U_p } \right) 
\quad . 
\label{thetaterm} 
\ee 
The $\theta$ term is a complicated function of the $z (x)$.  

For the auxiliary gauge field formulation 
in Eq.\ \bref{actionb}, 
a $\theta$ term can be defined by setting 
the link variable $U \left({  x, x+\Delta } \right)$ to be 
$\exp \left( { iA_{x,x+\Delta} } \right)$, 
letting $ U_p$ be the product of the 
link variables around a plaquette 
and using 
Eq.\ \bref{thetaterm}.  

At strong coupling, 
field configurations fluctuate considerably 
and might lead to bad behavior 
in Eq.\ \bref{thetaterm}.  
However, fluctuations are not so severe 
as to render 
Eq.\ \bref{thetaterm} so singular that 
strong-coupling computations 
cannot be performed.  
In fact, 
ref.\ \ct{seiberg84} 
found no computational difficulties in leading order. 
Likewise, 
no difficulties arose in the higher order calculations 
of the current work.  

Boundary conditions determine whether $Q$ 
is quantized.  
The analysis can be performed in a model-independent way.  
Since one is dealing with gauge-invariant functions 
of a compact $U(1)$ field, 
one can transform to a set 
of independent gauge-invariant degrees of freedom.  
In two-dimensions with open boundary conditions, 
the plaquette variables $U_p$ 
are such a set.  
Writing $U_p = \exp ( if_{ij} )$ 
with \mbox{$-\pi < f_{ij} \le \pi$}, 
Figure 1 shows a configuration of $U(1)$ 
fields which produces the most general 
configuration of the $U_p$.  
In this figure, 
most link variables equal $1$.  
A Wilson loop which runs around the boundary of the lattice 
has the value $\exp ( i \sum_{ij} f_{ij} )$.  
Since this factor is not necessary $1$, 
$Q$ need not be an integer.  

With periodic boundary conditions, 
the most general configuration 
is that of Figure 1 
with $\exp ( i \sum_{ij} f_{ij} ) = 1$ 
and for which 
all horizontal links are multiplied by 
$\exp ( i\alpha_1 )$ 
and all vertical links are multiplied by 
$\exp ( i\alpha_2 )$.  
These constant phases  
do not change the value of $U_p$ 
but allow the Polyakov lines 
at the lower-most and left-most sides of Figure 1 
to be non-zero.   
Since the condition $\exp ( i \sum_{ij} f_{ij} ) = 1$ 
implies that 
$ \sum_{ij} f_{ij} = 2 \pi Q$,  
$Q$ is integer.  
The constraint can be handled 
by inserting  
$ 
 \sum_Q \delta \left( { \sum_{ij} f_{ij}/(2\pi) - Q  } \right) 
$ 
into the functional integral. 
This makes the dependence on $\theta$ periodic 
since the following combination arises in the functional integral  
\be
  \exp \left( {i \frac{\theta}{2\pi} \sum_{ij} f_{ij} } \right) 
  \sum_Q \delta \left( { \sum_{ij} f_{ij}/(2\pi) - Q  } \right) = 
  \sum_m 
  \exp \left( {i \frac{\theta + 2 \pi m}{2\pi} \sum_{ij} 
  f_{ij} } \right) 
\quad . 
\label{ident} 
\ee
Here, the resummation formula 
$ 
  \sum_Q \delta \left( { \sum_{ij} f_{ij}/(2\pi) - Q  } \right) =
  \sum_m \exp \left( { i m \sum_{ij} f_{ij} } \right)
$ 
has been used.  

For the action in 
Eq.\ \bref{lactionb}, 
the change from link variables to plaquette variables $U_p$ 
can be explicitly performed with unit Jacobian.  
At $\beta = 0$, 
the integrand for the partition function 
$Z ( \beta=0 )$ is 
Eq.\ \bref{ident}, 
where the measure is 
$\prod_{ij} \int_{-\pi}^{\pi} df_{ij}/(2\pi) $: 
\be
  Z ( \beta ) \vert_{\beta=0} = 
    \sum_m \left[ { \frac{2}{\theta + 2 \pi m} 
     \sin \left( {\frac{\theta + 2 \pi m}{2}} \right)  } \right]^V
\quad , 
\label{z0} 
\ee 
where $V$, the number of sites of the lattice, 
is the volume of the system.  
The free energy ${\cal F}$ per unit volume, defined by 
$ {\cal F} = - \lim_{V \to \infty} ( \log Z )/ V$, 
is 
\be 
 {\cal F} =
  - \log \left( { \frac{2}{\theta}  
    \sin \left( {\frac{\theta}{2}} \right) 
  } \right) 
\quad , \quad {\rm for \ } -\pi < \theta \le \pi 
\quad , 
\label{f0} 
\ee 
and ${\cal F}$ is determined by periodicity in $\theta$ 
when $\theta$ is outside the range from $-\pi$ to $\pi$.
Eq.\ \bref{f0} 
is the result obtained in 
ref.\ \ct{seiberg84}.  
In what follows, 
we use periodic boundary conditions 
so that all quantities are periodic functions of $\theta$ 
with period $2 \pi$.  
Formulas below are written for the case 
$-\pi < \theta \le \pi$.   

To develop strong-coupling character expansions, 
we have found it calculationally more convenient 
to use the auxiliary gauge field formulation.  
Hence, 
throughout the rest of this work,  
we use the action which is the sum of  
Eqs.\ \bref{lactionb} and \bref{thetaterm}. 
The expansion procedure is similar to the one discussed in 
ref.\ \ct{samuel83}: 
One performs a group-theoretic-like Fourier decomposition 
of the action associated with each link.  
However, 
because of the $\theta$ term, 
one cannot immediately perform the integrations 
over the auxiliary gauge fields.  
One writes 
$$ 
  \exp  \beta N \left[{
    z_{x}^\ast \cdot z_{x+\Delta} 
    \exp \left( { iA_{x,x+\Delta} } \right) + 
    z_{x} \cdot z_{x+\Delta}^\ast 
    \exp \left( { -iA_{x,x+\Delta} } \right)  } \right] = 
$$ 
\be
  Z_0 (\beta) \, \sum_{ m, n} 
   d_{\left( { m ; n } \right) } \, 
   \exp \left[{  i (n-m) A_{x,x+\Delta} } \right] 
   z_{\left( { m ; n } \right) } ( \beta ) 
   \, f_{\left( { m ; n } \right) } 
   \left( { z_{x},  z_{x+\Delta} } \right)  
\quad , 
\label{expansion} 
\ee 
where 
$Z_0 (\beta)$ is
a normalization factor 
and where    
the sum is over non-negative integers $m$ and $n$.  
Character-like ``representations'' $r$ are governed by 
the pair $\left( { m ; n } \right)$: 
$r = \left( { m ; n } \right)$.  
Here, 
$d_{ r }$, 
$z_{ r }$ and
$f_{ r }$ 
are respectively 
character-like 
dimensions, 
expansion coefficients  
and representations 
for the $U(N)$ vector models 
\ct{samuel83}. 
See Appendix A 
for explicit formulas for low orders.  
 
With the appropriate normalization of the 
$f_{\left( {r } \right)}$, 
one has 
\be
  \int dv \, f_r (w,v)\,      f_{r^\prime}^\ast (w^\prime ,v) 
  = \frac{1}{d_r} \,  \delta_{r\,r^\prime}\, f_r(w,w^\prime) 
\quad , 
\label{ortho} 
\ee 
and 
\be
  f_r (w,w) = d_r
\quad . 
\label{dim} 
\ee 
The calculational procedure 
is to expand the action of each link as 
in Eq.\ \bref{expansion} 
and then perform the integrations over the $z (x)$ 
using 
Eq.\ \bref{ortho}.  
Lastly, 
one does the integration over the auxiliary gauge fields.  
Such integrations involve Wilson loops.  
Using Stoke's theorem, 
such Wilson loops can be written as products over plaquettes 
of plaquette variables to various powers.  
The result 
\be 
 { { 
       \int  dU_p \exp 
      \left[ {
       { {i \theta} \over {2\pi} }
       log \left( { U_p } \right) 
   }  \right]{ U_p}^n 
 } 
  \over 
  {  
         \int  dU_p \exp 
      \left[ {
       { {i \theta} \over {2\pi} }
       log \left( { U_p } \right) 
   }  \right]  
  } }
  = 
  (-1)^n \left( { {\theta} \over {\theta + 2\pi n} } \right) 
\quad 
\label{thetadep} 
\ee 
is useful.  

\bigskip 
{\bf\large \noindent III.\ Results}  
\setcounter{section}{3}   
\setcounter{equation}{0}   

It is straightforward to compute the partition function 
using the character-like expansion in 
Eq.\ \bref{expansion} 
and the above-mentioned procedure.  
Exponentiation of volume factors naturally occurs 
so that $ Z = \exp (- V {\cal F})$ 
where ${\cal F}$ is the free energy per unit volume.  
We have computed ${\cal F}$ to tenth order, 
which is four more terms than in 
ref.\ \ct{seiberg84}.  
When expanded in $\beta$ up to order $\beta^2$, 
our results agree with those of 
ref.\ \ct{seiberg84}.  
The fourth, sixth, eighth and tenth orders 
provide new information 
about the $CP^{N-1}$ models in the presence 
of a $\theta$ term.  
Because the result for ${\cal F}$ 
is rather lengthy, 
we have relegated it 
to Appendix B.  
See Eq.\ \bref{freeenergy}. 

An analysis of the strong-coupling expansion leads to 
some immediate physical results.  
The first is the loss of superconfinement 
when $\theta \ne 0$.  
When $\theta \ne 0$, 
strong-coupling diagrams appear in fourth  
and higher order terms 
in which a positively charged loop does not have to be directly 
on top of a negatively charged loop.  
Local charge liberation occurs.  
Charges are no longer superconfined.  
Instead they are bound together by ordinary confinement.  
Thus, the confining force is weakened.  
Furthermore, 
charged loops oriented counterclockwise 
arise with probabilities 
which are different from loops oriented clockwise.  
This is expected 
and is an indication of the explicit CP violation 
generated by a $\theta$ term 
when $\theta \ne 0$ or $\theta \ne \pi$.  

For $\beta$ small, 
higher order terms do not significantly change 
the result in
Eq.\ \bref{f0}:   
${\cal F}$ increases as $\vert \theta \vert $ increases 
and its graph is concave upward.  
As $\beta$ becomes larger, 
${\cal F}$ is lowered 
particularly for larger values of $\theta$.  
Then, for sufficiently large $\beta$ 
a maximum in ${\cal F}$ arises at $\theta$ 
near $\pm \pi$.  
Figure 2a  
illustrates this for the $N=4$ case.  
Other values of $N$ are qualitatively similar.  
As $\beta$ is further increased, 
the maximum in ${\cal F}$ occurs at smaller values 
of $\theta$.  
See Figure 2b which plots ${\cal F}$ versus $\theta$ 
when $\beta = 0.7$.  
For the $N=4$ case, 
the maximum eventually reaches $\theta = 0$ 
for $\beta \approx 0.90$.  
However, 
at these moderate values of $\beta$,  
higher order terms might be important.  
  
In two dimensions 
the string tension $\sigma (e, \theta, \beta)$ 
between a charge $e$ 
and a charge $-e$ 
can be obtained from the free energy 
via the standard formula 
$ 
 \sigma (e, \theta , \beta ) = 
   {\cal F} (\theta + 2 \pi e , \beta) - {\cal F} (\theta , \beta ) 
$.\ct{luscher78}  
Hence, 
the value of $\theta$ 
at which the free energy stops increasing 
and begins to decrease 
marks the point where confinement is lost 
for infinitesimal charges.  
Using our results for ${\cal F}$, 
a phase diagram can be constructed.  
It is shown in Figure 3 for the $CP^3$ model.  
The reader should be forewarned 
that this phase diagram  
is obtained under the assumption that strong coupling results 
can be extrapolated to moderate values of $\beta$. 
The dotted line in Figure 3 indicates where CP invariance 
spontaneously breaks.  
This line bifurcates at around $\beta = 0.56$.  
For $\beta$ larger than $0.56$ there are two phases 
as a function of $\theta$. 
The region labeled ``infinitesimal confinement loss'' 
is where charges of very small magnitude 
no longer experience a confining force.  
The phase diagram of Figure 3, 
which is obtained by analytic methods, 
is qualitatively similar 
to the diagram of 
refs.\ \ct{schierholz94}, 
obtained from Monte Carlo simulations 
of the adjoint form of the $CP^3$ action in 
Eq.\ \bref{lactiona}.  
We have also taken our tenth order series, 
expanded it in $\beta$, and obtained Pad\'e approximates 
to try to increase the range of validity of our 
strong-coupling results.  
For the diagonal approximate, 
the $\beta$ at which infinitesimal confinement loss
first takes place is shifted to larger values. 
Even for $\beta$ as large as $2.0$, 
there is confinement for $\theta$ small.  
Hence, the phase diagram using the diagonal Pad\'e approximate 
looks as in Figure 3 but with the lines separating 
the regions 
``full confinement'' from ``infinitesimal confinement loss'' 
shifted to the right.

The mass gap $m$ can be extracted from the two-point function: 
\be 
  \sum_x \langle { 
  z(0,0) \cdot z^\ast (x,L) \,  z^\ast (0,0) \cdot z(x,L) 
  } \rangle \sim \exp (-m L) 
\quad 
\label{cor} 
\ee   
Summing over $x$ to project onto zero momentum makes it easier 
to compute $m$. 
The strong-coupling series for $m$  
at $\theta = 0$ has been obtained to tenth order 
in ref.\ \ct{rs81}. 
The computation for $\theta \ne 0$ is considerably more complicated 
in that sets of graphs must be summed. 
Here, we compute $m$ to order four when a $\theta$ term is present.  
This is one more order than the $\beta$ expansion computation 
in ref.\ \ct{seiberg84}.   
Our results agree with 
\ct{rs81} and \ct{seiberg84} 
for those terms which are in common.  
The computation of $m$ is lengthy and 
is relegated to Appendix B.  
See Eq.\ \bref{massgap}.  

The second order $\theta$-dependent term 
lowers the value of $m$.  
This is physically expected.  
The loss of superconfinement 
means that the charges can move more freely 
within the bound state, 
thereby lowering their constituent mass.  
However, the fourth order 
$\theta$-dependent term 
counteracts the effect 
and tends to increase the value of $m$ 
particularly for large $\theta$.  
In Figure 4, 
we plot $m$ versus $\beta$ for various fixed values 
of $\theta$.  
For sufficiently small $\beta$, 
$m$ is lowered when $\theta$ is non-zero, 
but the effect is small and not too visible 
for the scales used in Figure 4.  
The curves for $\theta \le 1.0$ 
are almost identical to the case $\theta = 0$ 
and are displayed as one curve.  
For $\theta > 1.0$, 
fourth order contributions increase $m$ significantly 
at moderate values of $\beta$, 
however, at such $\beta$ values, 
higher order terms can be important.  
 
For sufficient large $\beta$, 
refs.\ \ct{schierholz94} 
has observed a phase transition 
as $\theta$ is increased.  
A priori such a transition 
could be due to the vanishing of $m$.  
Our computation of $m$ suggests that 
this is not the case 
and supports the conclusion 
that the transition is due to the loss of confinement.  
If confinement is lost, 
one expects on physical grounds that 
$m = 2 m_0 > 0$ 
where $m_0$ is the mass of a free charged particle.  
  
{\bf\large \noindent IV.\ Conclusions}  
\setcounter{section}{4}   
\setcounter{equation}{0}   

In this work, 
we have developed and performed 
the first character expansions 
for a lattice model with a $\theta$ term.  
The purpose was to gain insight into the 
strong-coupling behavior of the physics 
associated with a topological term.  

We found the following physical effects. 
The superconfinement property of the $CP^{N-1}$ 
is lost for $\theta \ne 0$.  
When $\theta \ne 0$, 
superconfinement liberation occurs.  
A bound state still persists but is tied together 
by ordinary confining forces.  
The weakening of the binding in the bound state 
is reflected by the lowering of its mass 
at small $\beta$.  

In Monte Carlo simulations, 
refs.\ \ct{schierholz94} observed a phase transition 
when $\theta$ was increased.  
It was argued that this transition 
was associated with the loss of confinement.  
Our analytic computations, although not conclusive 
and performed with a different lattice version 
of the $CP^{N-1}$ model, 
support this conclusion.  
Furthermore, 
the mass of the bound state 
does not appear to vanish at the deconfinement point.

The main motivation for our work 
is to gain insight into the strong CP problem.  
Our analytic computations support the conclusion 
of refs.\ \ct{schierholz94} 
that the analog of the strong CP problem 
is naturally resolved in the $CP^{N-1}$ models. 
It was argued that, as $\beta \to 0$, 
$\theta$ must be adjusted to $0$ 
as the continuum limit is taken 
in order to have a confining theory with bound states.  
Our analytic computations support this statement. 
Assuming that higher-order corrections do not invalidate results, 
the strong coupling computations  
show that  
$\theta$ must be decreased as $\beta$ is increased 
to remain in the confining phase, 
but they are unable to say whether $\theta$ 
must be taken to zero as $\beta \to \infty$.

The conclusion concerning the strong CP problem 
appears rather specific to two dimensions.  
In two dimensions in the continuum, 
a $\theta$ term in the action 
produces an immediately physical effect -- 
it introduces charges at spatial infinity 
which create a constant background electric field. 
This has a direct effect on a bound state.  
In four-dimensional gauge theories, 
a $\theta$ term affects physics more indirectly:   
If a theory possesses monopoles, 
then such monopoles acquire an electric charge 
proportional to $\theta$ 
\ct{witten79c}.  
In Monte Carlo simulations 
at a particular value of the coupling, 
ref.\ \ct{schierholz95} has shown that the $SU(2)$ 
four-dimensional Yang-Mills theory  
has a dependence on $\theta$ similar to 
that of the $CP^{N-1}$ models. 
In particular, 
a phase transition occurs at a certain value of $\theta$.  
Given the difference between two and four dimensions, 
it is not probable that this transition 
is related to deconfinement.  
It is thus quite important to determine 
the physics associated 
with the transition 
to decide whether $\theta$ must be tuned to zero 
as the continuum limit is taken.

\medskip 
{\bf\large\noindent Acknowledgments}  

S.\ Samuel thanks Columbia University for hospitality and support.   
This work was supported in part 
by the Humboldt Foundation under a Lynen-Fellowship and 
by the National Science Foundation under the grant  
(PHY-9420615).

\medskip

%
%

\let\a=\alpha    \let\b=\beta     \let\g=\gamma    \let\d=\delta
     \let\e=\epsilon
\let\z=\zeta      \let\h=\eta      \let\q=\theta       \let\i=\iota
 \let\k=\kappa
\let\l=\lambda  \let\m=\mu      \let\n=\nu            \let\x=\xi
 \let\p=\pi
\let\r=\rho        \let\s=\sigma  \let\t=\tau            \let\u=\upsilon
   \let\f=\phi
\let\c=\chi        \let\y=\psi       \let\vq=\vartheta       \let\vf=\varphi
\let\w=\omega

\let\X=\Xi          \let\P=\Pi        \let\S=\Sigma      \let\U=\Upsilon
 \let\F=\Phi
\let\Y=\Psi        \let\W=\Omega      \let\D=\Delta     \let\L=\Lambda

\let\la=\label    \let\nn=\nonumber     \let\fr=\frac        \let\ov=\over
\let\bl=\bigl      \let\br=\bigr              \let\Bl=\Bigl
    \let\Br=\Bigr

\let\na=\nabla      \let\pa=\partial      \let\bm=\bibitem 
\newcommand{\bel}{\begin{equation}}
\newcommand{\eel}[1]{\label{#1}\end{equation}}
\newcommand{\bea}{\begin{eqnarray}}
\newcommand{\eea}{\end{eqnarray}}
\newcommand{\ra}{\rightarrow}
\newcommand{\longra}{\longrightarrow}
\newcommand{\Math}[1]{\mbox{$#1$}}


\newcommand{\sect}[1]{\setcounter{equation}{0} \section{#1}}
\renewcommand{\theequation}{\thesection .\arabic{equation}}
\newcommand{\refer}[1]{(\ref{#1})}

\bigskip 
\renewcommand{\theequation}{A.\arabic{equation}}
\setcounter{equation}{0}
{\bf\large\noindent Appendix A: Characterlike Expansion Quantities}

The leading order character-like representations 
$f_r$ used in our computation are
\bea
f_{(0;0)}(v,w)&=&1 \ , \qquad \qquad 
f_{(1;0)}(v,w)=\sqrt{N}\, v\cdot w \ , \nn\\
f_{(2;0)}(v,w)&=& \sqrt{\frac{N(N+1)}{2}} \, 
  (v^\ast\cdot w)^2 \ , \nn\\
f_{(1;1)}(v,w)&=& N\sqrt{\frac{N+1}{N-1}}\Bigl [ v^\ast\cdot w\,
 v\cdot w^\ast - \frac{1}{N}\Bigl ] \ , \la{frs} \\
 f_{(2;1)}(v,w)&=& (N+1) 
    \sqrt{\frac{N(N+2)}{2(N-1)}}\Bigl [ v^\ast\cdot w 
\, (v\cdot w^\ast)^2 - \frac{2}{N+1} 
   v\cdot w^\ast\Bigl ] \ , \nn \\
 f_{(m;l)}^\ast(v,w)&\equiv&f_{(l;m)}(v,w) 
     \equiv f_{(m;l)}(w,v) \quad ,\nn
\eea
and the corresponding dimensions $d_r$ are
\bea
d_{(0;0)}&=&1 \ , 
  \qquad d_{(1;0)}=\sqrt{N} \ , \qquad d_{(2;0)}=
\sqrt{\frac{N(N+1)}{2}} \ , \nn\\
d_{(1;1)}&=&\sqrt{N^2-1} \ , \qquad d_{(2;1)} = 
  \sqrt{\frac{N(N-1)(N+2)}{2}} \ , 
\la{drs}\\ 
d_{(l;m)}&=&d_{(m;l)} 
\quad . \nn
\eea
These formulas can be derived with the help of the integral
\bel
\int dv \, v_{i_1}\, v_{i_2}\ldots v_{i_m}\, v^\ast_{j_1}\, v^\ast_{j_2}
\ldots v^\ast_{j_m}= \frac{(N-1)!}{(N-1+m)!}\sum_{\s\in S_m} 
\d^{i_1}_{j_{\s(1)}}\, \d^{i_2}_{j_{\s(2)}}\ldots \d^{i_m}_{j_{\s(m)}} \ ,
\eel{deltas}
where the sum is to be taken over all permutations of the $j$ indices. 
In Eq.\ \refer{deltas} 
and throughout this appendix we are using the following 
normalized notation for vector integrals
\bel
\int dv\, (\ldots)= \frac{\int dv \, \d(v^\ast \cdot v-1)\, (\ldots)}
{\int dv \, \d(v^\ast \cdot v-1)} \quad .
\eel{vectorintegrals} 

The expansion coefficients \Math{z_r(\b)} may be computed by inverting
eq.\  \refer{expansion} using eq.\ \refer{ortho}:
\bel 
z_{(l;m)}(\b)= \frac{1}{d_{(l;m)} 
 \, Z_0(\b)}\int dv f_{(l;m)}(v,w)\, \exp [
N\, \b\, ( v\cdot w^\ast + v^\ast \cdot w)] 
\quad .
\eel{zU(N)}
The \Math{z_{(l;m)}(\b)} are actually independent of $w$ and are of order
\Math{l+m}, i.e. \Math{z_{(l;m)}= O(\b^{l+m})}. 

The factor \Math{Z_0(\b)} normalizes \Math{z_{(0;0)}} to $1$ and
can be explicitly computed
\bel 
Z_0(\b)= \frac{(N-1)!\, I_{N-1}(2\, \b\, N)}{(\b\, N)^{N-1}} 
\quad ,
\eel{F0compute}
where \Math{I_n(x)} is the Bessel function \Math{I_n(x)=\int_{-\p}^\p
d\q/(2\p)\, \exp[in\q + x \cos \q]}.

The \Math{z_{(l;m)}(\b)} can also be explicitly computed and read
\bea
z_{(0,0)}&=& 1 \ ,  \nn \\
z_{(1;0)}&= &\frac{I_N(2\, \b\, N)}{I_{N-1}(2\, \b\, N)}\nn\\
 &= &\b- \frac{N}{N+1}\b^3 + 
  \frac{2\, N^2}{(N+1)(N+2)} \b^5  \nn \\  
 & \ & - {{{{\beta}^7} {N^3} \left( 5 N + 6  \right) }\over 
     {{{\left( N + 1 \right) }^2} \left( N + 2 \right)  
       \left( N + 3 \right) }} + 
  {{2 {{\beta}^9} {N^4} \left( 7 N +12  \right) }\over 
    {{{\left( N + 1 \right) }^2} \left( N + 2 \right)  
      \left( N + 3 \right)  \left( N + 4 \right) }} \nn \\ 
  & \ & - {{2 {{\beta}^{11}} {N^5} 
      \left( 21 {N^3}  + 118 {N^2} + 214 N + 120  \right) }\over 
    {{{\left( N + 1 \right) }^3} {{\left( N + 2 \right) }^2} 
      \left( N + 3 \right)  \left( N + 4 \right)  
      \left( N + 5 \right) }} +  \ldots 
  \ , \nn \\
z_{(2;0)}&=&z_{(1;1)}= 1- \frac{1}{\b}\,\, z_{(1;0)}  \ , \\
z_{(2;1)}&=&-\frac{N+1}{\b\, N}+ \Bigl ( 1 + \frac{N+1}{\b^2\, N} \Bigl ) \,\,
  z_{(1;1)} \quad . \nn
\eea 
Here, we have displayed $z_{(1;0)}$ to order eleven.  
From this expansion, 
one can construct the ordinary $\beta$ expansions 
of the free energy and mass gap.  

After the expansion of each link as 
in eq.\ \refer{expansion}, 
one has to
perform the integral over the vectors and the link angles. 
For the vectors 
it is convenient to make use of the relations
\bea
f_{(1;0)}\, f_{(1;0)} &=& \sqrt{\frac{2N}{(N+1)}}\, f_{(2;0)} \ , \nn \\
f_{(1;0)}\, f_{(0;1)} &=& \sqrt{\frac{N-1}{N+1}}\, f_{(1;1)} + 
   f_{(0;0)} \ , \nn \\
f_{(1;0)}\, f_{(1;1)} &=& N\sqrt{\frac{2}{(N+1)(N+2)}}\, f_{(2;1)} + 
 \sqrt{\frac{N-1}{N+1}}\, f_{(1;0)} \ , \la{CGrel} \\
f_{(0;1)}\, f_{(2;1)} &=& N\sqrt{\frac{2}{(N+2)(N+3)}}\, f_{(2;2)} +
  N\sqrt{\frac{2}{(N+1)(N+2)}}\, f_{(1;1)} 
\quad .\nn
\eea

\bigskip 
\renewcommand{\theequation}{B.\arabic{equation}}
\setcounter{equation}{0}
{\bf\large\noindent Appendix B:  
Results for the Free Energy and the Mass Gap}  

We have computed the free energy 
to order 10 and the mass gap to order
4 in an \Math{\q}-exact strong-coupling character expansion.
The result for the free energy ${\cal F}$ is
\bea
  \lefteqn{ - {\cal F} = \frac{1}{V}\ln {\cal Z} = } \nn \\
& &
 2 \ln Z_0(\beta)+ 
 \ln [\frac{2}{\theta} \sin {{\theta }\over 2}] + 
 {{d_{(1;0)}}^2} \left( {{\theta }\over {2 \pi  - \theta }} - 
     {{\theta }\over {2 \pi  + \theta }} \right)  {{z_{(1;0)}}^4}\nn \\ & & 
         + 
  2 {{d_{(1;0)}}^2} \left( {{{{\theta }^2}}\over 
       {{{\left( 2 \pi  - \theta  \right) }^2}}} + 
     {{{{\theta }^2}}\over {{{\left( 2 \pi  + \theta  \right) }^2}}}
      \right)  
   {{z_{(1;0)}}^6}                                 \nn \\ & & 
 + {d_{(1;1)}}^2 {z_{(1;1)}}^4
 + 5 {d_{(1;0)}}^4 {{{\theta }^2} 
   \over 
    {\left( 2 \pi  - \theta  \right)  \left( 2 \pi  + \theta  \right) 
    }}{z_{(1;0)}}^8+                                      \nn \\ & & 
  {d_{(1;0)}}^2 \left( {{{\theta }^4}\over 
       {{\left( 2 \pi  - \theta  \right) }^4}} + 
     {{{\theta }^4}\over {{\left( 2 \pi  + \theta  \right) }^4} }
     \right)  {z_{(1;0)}}^8 \nn \\ & & +          
    6 {{d_{(1;0)}}^2} 
   \left( {{{{\theta }^3}}\over {{{\left( 2 \pi  - \theta  \right) }^3}}} - 
     {{{{\theta }^3}}\over {{{\left( 2 \pi  + \theta  \right) }^3}}} 
     \right)  
   {{z_{(1;0)}}^8} \nn \\ & & 
    - {5\over 2} {d_{(1;0)}}^4
      \left( \frac{{\theta }^2}{( 2 \pi  - \theta  ) ^2} + 
       \frac{{\theta }^2}{( 2 \pi  + \theta )^2} \right )  
        {z_{(1;0)}}^8  \nn \\ & &  + 
        {2 {\sqrt{{N-1}\over {N+1}}} {{d_{(1;0)}}^2} d_{(1;1)} 
      \left( {{{{\theta }^2}}\over 
          {{{\left( 2 \pi  - \theta  \right) }^2}}} + 
        {{{{\theta }^2}}\over {{{\left( 2 \pi  + \theta  \right) }^2}}} \right
        )  {{z_{(1;0)}}^6} z_{(1;1)}}\nn \\ & & 
         - 
  4 \sqrt{{2N}\over{N + 1}} {d_{(1;0)}}^2 d_{(2;0)} {{{\theta }^2} \over
  {( 2 \pi  - \theta)( 2 \pi  + \theta  )}  }
     {z_{(1;0)}}^6 z_{(2;0)} \nn \\ & &
      + {d_{(2;0)}}^2 \left( {{\theta }\over {\theta -4 \pi  }} + 
     {{\theta }\over {\theta +4\pi }} \right)  {z_{(2;0)}}^4  \nn \\ & & + 
  2 {{d_{(1;0)}}^2} \left( {{{{\theta }^6}}\over 
       {{{\left( 2 \pi  - \theta  \right) }^6}}} + 
     {{{{\theta }^6}}\over {{{\left( 2 \pi  + \theta  \right) }^6}}} 
     \right)  
   {{z_{(1;0)}}^{10}} \nn \\ & &
    + 
    8 {{d_{(1;0)}}^2} 
   \left( {{{{\theta }^5}}\over {{{\left( 2 \pi  - \theta  \right) }^5}}}  - 
     {{{{\theta }^5}}\over {{{\left( 2 \pi  + \theta  \right) }^5}}} 
     \right)  
   {{z_{(1;0)}}^{10}} \nn \\ & &  + 
   18 {{d_{(1;0)}}^2} 
   \left( {{{{\theta }^4}}\over {{{\left( 2 \pi  - \theta  \right) }^4}}} + 
     {{{{\theta }^4}}\over {{{\left( 2 \pi  + \theta  \right) }^4}}}
      \right)  
   {{z_{(1;0)}}^{10}} 
   \nn
    \\ & &- 16 {{d_{(1;0)}}^4} 
   \left({\theta \over {2 \pi  - \theta }}  - 
     {\theta  \over {2 \pi  + \theta }}  \right )
   \left( {{{{\theta }^2}}\over 
             {{{\left( 2 \pi  - \theta  \right) }^2}}} + 
           {{{{\theta }^2}}\over 
           {{{\left( 2 \pi  + \theta  \right) }^2}}} 
           \right) {{z_{(1;0)}}^{10}} \nn \\ & &
   + 
  {{12 \sqrt{{N-1}\over {N+1}} {{d_{(1;0)}}^2} d_{(1;1)} 
      \left( {{{{\theta }^3}}\over 
          {{{\left( 2 \pi  - \theta  \right) }^3}}} - 
        {{{{\theta }^3}}\over {{{\left( 2 \pi  + \theta  \right) }^3}}} \right
        )  {{z_{(1;0)}}^8} z_{(1;1)}}}\nn \\ & &
     + 
  4 \sqrt{{N-1}\over{N + 1}} {d_{(1;0)}}^2 d_{(1;1)}
      \left( {{\theta }\over {2 \pi  - \theta }} - 
        {{\theta }\over {2 \pi  + \theta }} \right)  {z_{(1;0)}}^4
      {z_{(1;1)}}^3\la{freeenergy} \\ & &
     + 
  12 \sqrt{{2N}\over{N + 1}}  {d_{(1;0)}}^2 d_{(2;0)} 
      \left( {{{\theta }^3}\over 
          {\left( 2 \pi  - \theta  \right)  
            \left( 2 \pi  + \theta  \right)^2 }}- 
        {{{\theta }^3}\over 
          {\left( 2 \pi  - \theta  \right) ^2
            \left( 2 \pi  + \theta  \right) }}\right)  {z_{(1;0)}}^8
      z_{(2;0)}\nn \\ & &
    + 
  4 \sqrt{{2N}\over{N+1}} {d_{(1;0)}}^2 d_{(2;0)} 
      \left( {{{\theta }^2}\over 
          {\left( 2 \pi  - \theta  \right)  \left( \theta  -4 \pi  
          \right) }
          } - {{{\theta }^2}\over 
          {\left( 2 \pi  + \theta  \right)  \left( \theta +4 \pi  
           \right) }}\right)  {z_{(1;0)}}^4 {z_{(2;0)}}^3 
\  . 
\nn
 \eea
For the mass gap up to order 4 in $\b$, 
we obtained the result
 \bea
 \lefteqn{a\cdot m(\q,\b)=} \nn \\
 && -\ln {z_{(1;1)}} - 2\, {z_{(1;1)}} - 4  \frac{N}{N+1} {z_{(1;0)}}^2 \left
  ( \frac{  \k \, a}{1-  \k \, a}+ \frac{  \k \, b}
 {1-  \k \, b}\right ) \nn \\
 && - \frac{N}{N+1} {z_{(1;0)}}^4 \Biggl[ 9 \left( \frac{  \k \, a^2}{1-  \k 
 \, a^2} + 
 \frac{  \k \, b^2}{1-  \k \, b^2}
 \right ) + 24 \Biggl ( \frac{a (   \k \, a )^2}{(1-  \k \, a)(1-  \k \, a^2)} \nn 
 \\ &&+
 \frac{b (   \k \, b )^2}{(1-  \k \, b)(1-  \k \, b^2)}\Biggr)
  + 16 \left
  ( \frac{a (  \k \, a)^3}{(1-  \k \, a^2)
 (1-   \k \, a)^2}+ \frac{b (  \k \, b)^3}{(1-  \k \, b^2) (1-   \k \, b)^2} \right) 
 \Biggr ] \nn \\ 
 && 
 +16 \left( \frac{N}{N+1}\right )^2 \, {z_{(1;0)}}^4\, \left ( \frac{  \k \, a}
 {1-  \k \, a}
 +\frac{  \k \, b}{1-  \k \, b}
 \right )\,
 \left( \frac{  \k \, a}{(1-  \k \, a)^2}+\frac{  \k \, b}{(1-  \k \, b)^2}\right ) 
 \nn \\
 && 
 +4 \frac{N}{(N+1)^2}\, {z_{(1;0)}}^2\, {z_{(1;1)}}\, \left
  [\frac{(  \k \, a)^2}{(1-  \k \, a)^2}+
 \frac{(  \k \, b)^2}{(1-  \k \, b)^2} \right ]\nn \\
 && 
 -8 \left ( \frac{N}{N+1}\right )^2 \,  {z_{(1;0)}}^2\, {z_{(2;0)}}\, 
 \frac{(  \k \, a)(  \k \, b)}{(1-  \k \, a)(1-  \k \, b)} \nn \\ & &
 +8  \frac{N}{N+1} \, {z_{(1;0)}}^2\,
 {z_{(1;1)}}\, \left [ \frac{   \k \, a}{(1-   \k \, a)^2}+\frac{  \k \, b}{(1-  
 \k \, b)^2} 
 \right ] \nn \\
 && 
- 2\, \frac{N^2}{N+1}\, {z_{(1;0)}}^3\, \frac{z_{(2;1)}}{{z_{(1;1)}}}\, 
(a+b) + 2\, N {z_{(1;0)}}^4\,
 ( a+b) 
\quad ,
\la{massgap}
\eea
where
 \bel
  \k =\frac{{z_{(1;0)}}^2}{z_{(1;1)}} \ , 
\qquad a=\frac{\q}{2\p-\q} \ , \qquad
  \mbox{and}\qquad
 b=-\frac{\q}{2\p+\q} 
\quad .
\eel{kappaab}
The above result is restricted to the $\b$-dependent domain of 
\bel
 |\q | < 2\p\, \frac{z_{(1;1)}}{z_{(1;1)}+{z_{(1;0)}}^2 }\approx 2\p\, 
 \frac{N}{2N+1}\, (1+ O(\b)) 
\quad .
\eel{domain}

\pagebreak

\def\NPB#1#2#3{ {Nucl.{\,}Phys.{\,}}{\bf B{#1}} ({#3}) {#2}} 
\def\PLB#1#2#3{ {Phys.{\,}Lett.{\,}}{\bf {#1}B} ({#3}) {#2}} 
\def\PRL#1#2#3{ {Phys.{\,}Rev.{\,}Lett.{\,}}{\bf  {#1}} ({#3}) {#2}} 
\def\PRD#1#2#3{ {Phys.{\,}Rev.{\,}}{\bf D{#1}} ({#3}) {#2}} 
\def\PR#1#2#3{ {Phys.{\,}Rep.{\,}}{\bf {#1}} ({#3}) {#2}} 
\def\OPR#1#2#3{ {Phys.{\,}Rev.{\,}}{\bf {#1}} ({#3}) {#2}} 
\def\NC#1#2#3{ {Nuovo Cimento{\,}}{\bf {#1}} ({#3}) {#2}}


\bigskip   
{\bf\large\noindent Figure Captions}  
\medskip 

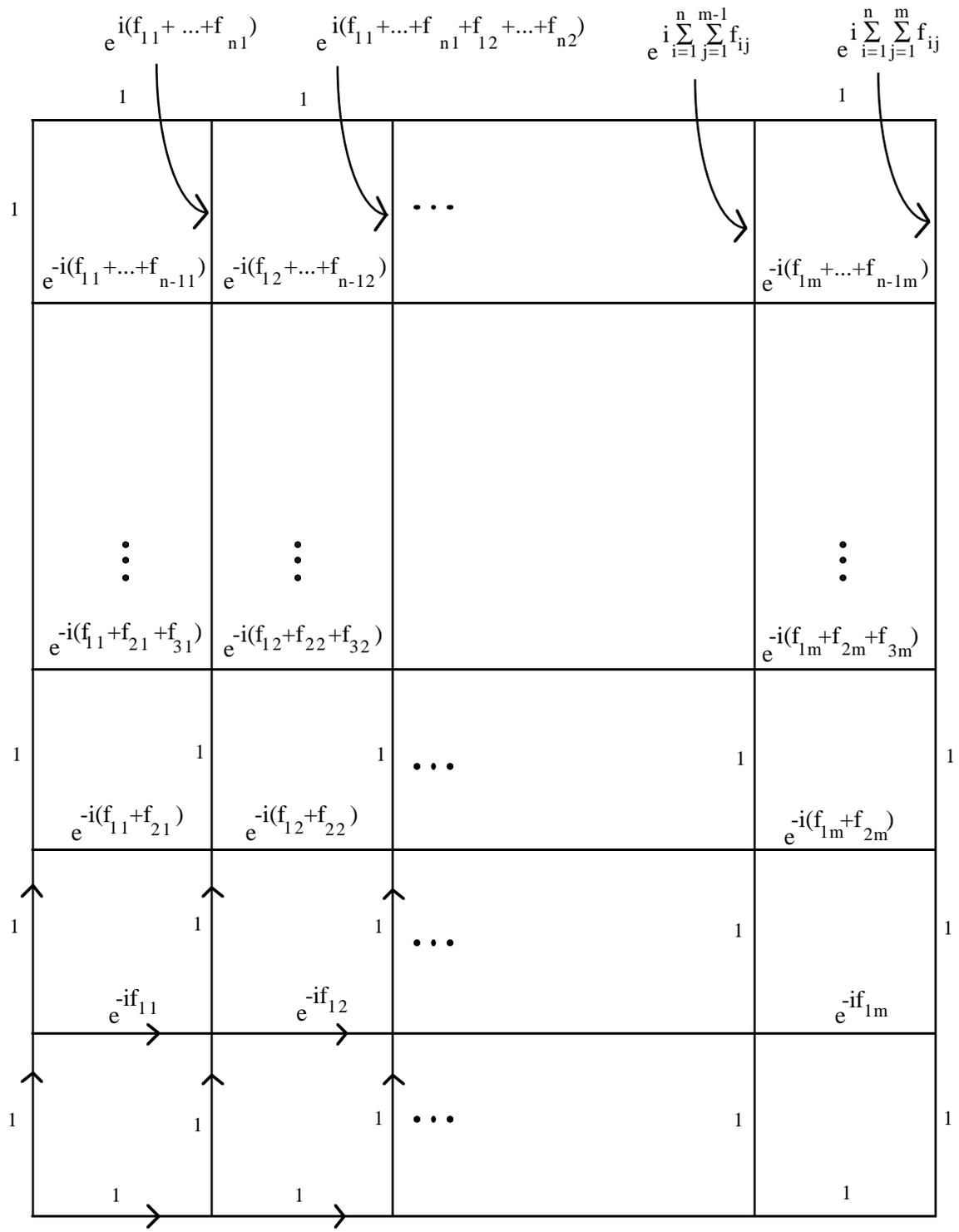
Figure 1. A Lattice Link Configuration which Produces the Most 
General Plaquette-Variable Configuration 
for Open Boundary Conditions.    

\medskip 

Figure 2a. The Free Energy Per Unit Volume ${\cal F}$ 
as a Function of $\theta$. 
Shown here is the order-by-order case 
of $CP^{3}$ with $\beta = 0.6$.  

\medskip 

Figure 2b. The Free Energy Per Unit Volume ${\cal F}$ 
as a Function of $\theta$. 
Shown here is the case of $CP^{3}$ with $\beta = 0.7$.  
  
\medskip 

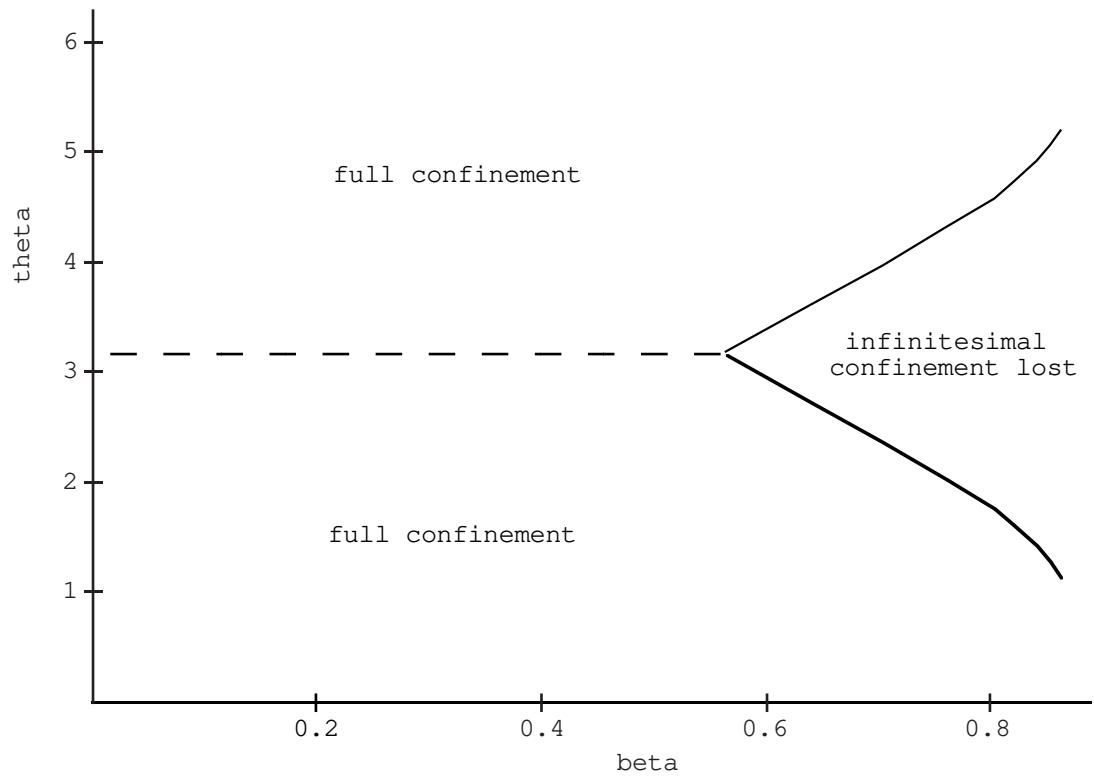
Figure 3. The Phase Diagram for the $CP^{3}$ Model 
as Determined by the Tenth Order Free Energy Result.  
  
\medskip 

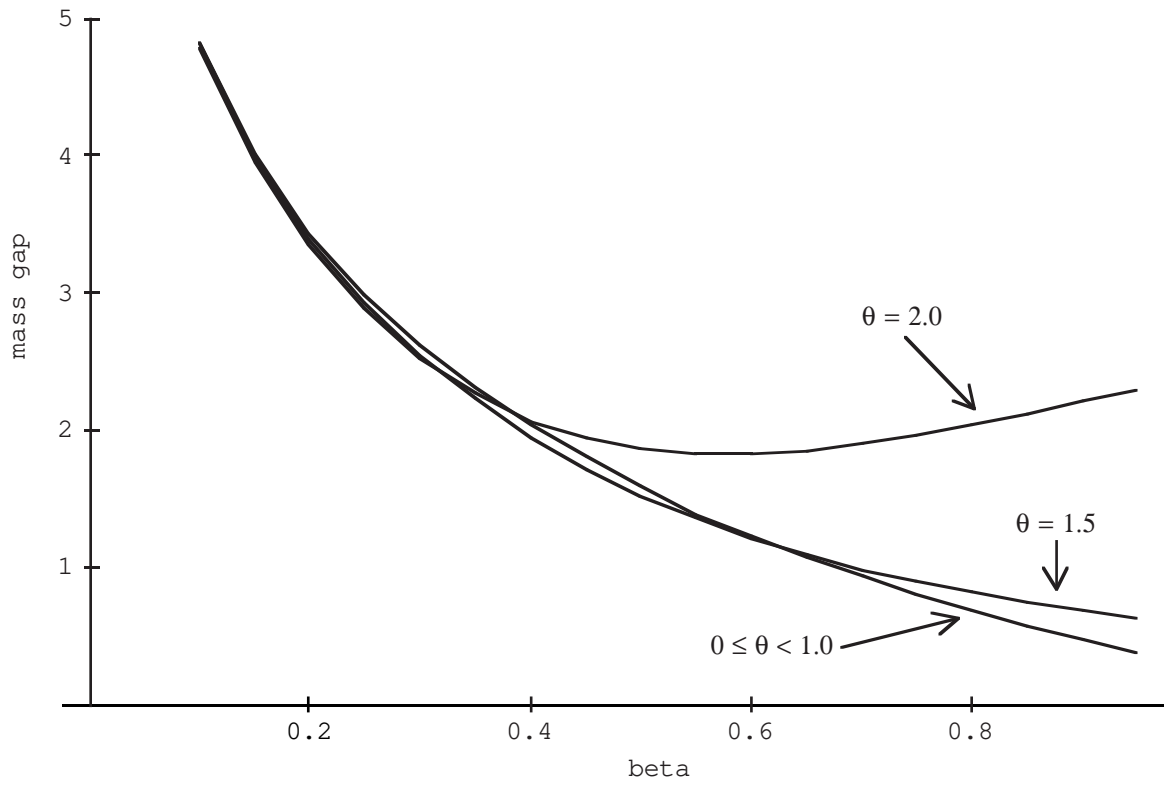
Figure 4. The Fourth-Order Result for the Mass Gap $a m$ 
as a Function of $\beta$ 
for Various Values of $\theta$.

\vfill\eject 
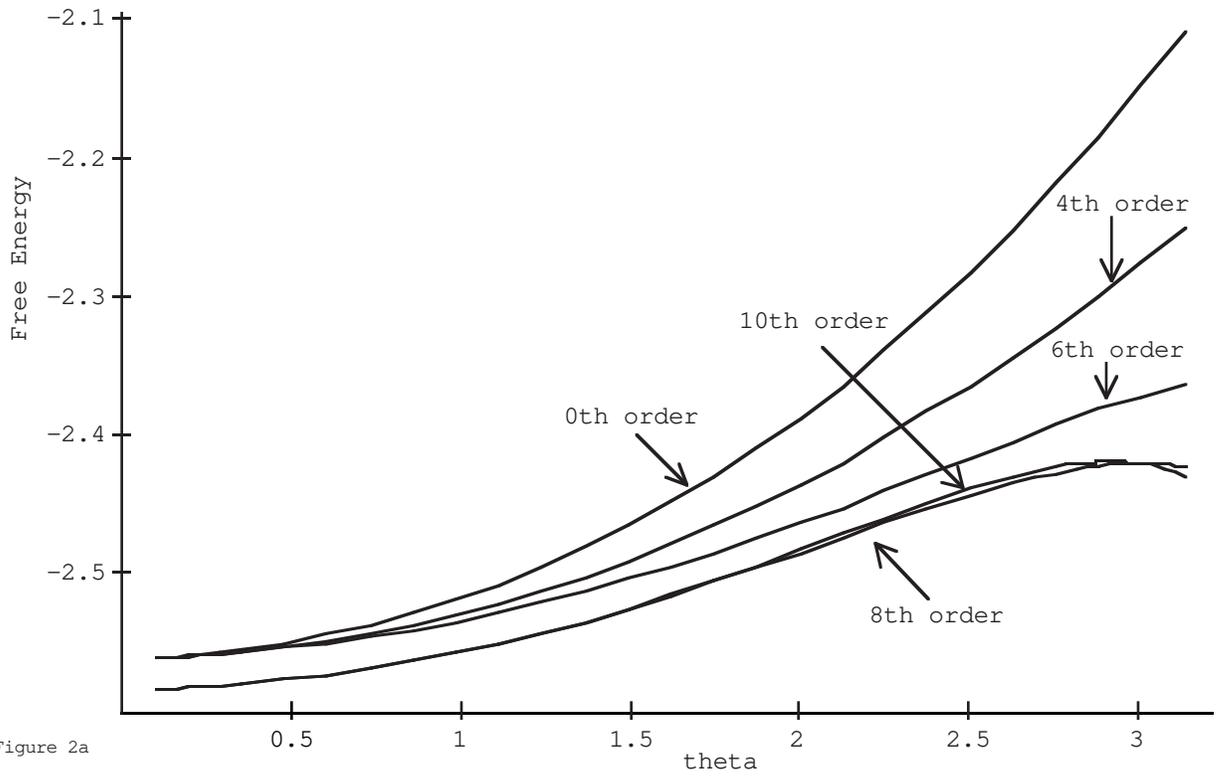
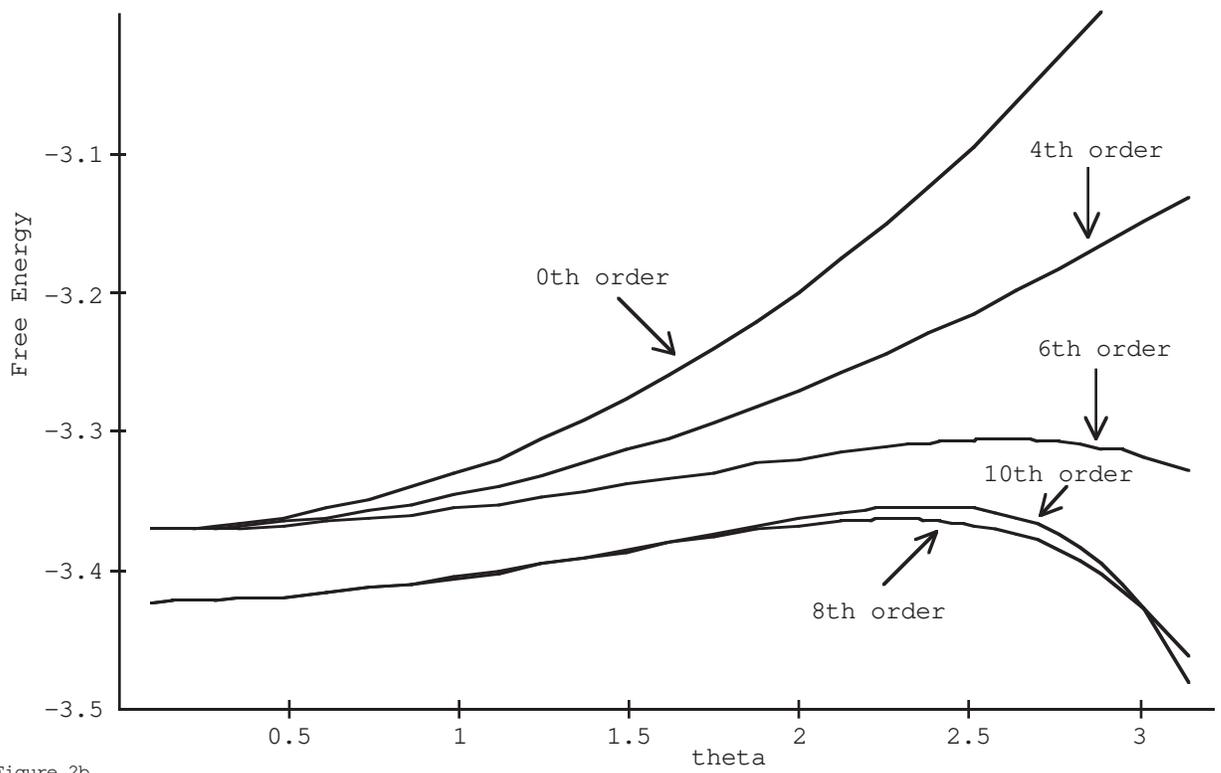
\end{document}